\pdfoutput=1
\PassOptionsToPackage{square,sort,comma,numbers}{natbib}
\PassOptionsToPackage{usenames,dvipsnames,pdftex}{xcolor}

\documentclass[amsfonts,amssymb,10pt, a4paper, final, nofootinbib, longbibliography, twocolumn,notitlepage,hyperref,twoside,allowtoday,noarxiv, accepted=2019-08-09]{quantumarticle}

\usepackage{amsmath,amsfonts,amssymb,amsthm,mathtools,
	IEEEtrantools,bm,graphicx,verbatim,mathrsfs,units}

\usepackage[raggedright,bf,SF]{subfigure}
\usepackage[square,sort,comma,numbers]{natbib}
\usepackage[usenames,dvipsnames,table,rgb]{xcolor}
\definecolor{blue}{rgb}{0,0,1}
\definecolor{grey}{rgb}{0.6,0.6,0.6}

\usepackage{placeins} 
\usepackage{tikz}
\usetikzlibrary{calc,decorations.pathreplacing}
\usetikzlibrary{arrows,shapes}
\usetikzlibrary{calc,decorations.pathreplacing}

\definecolor{myurlcolor}{rgb}{0,0,0.7}
\definecolor{myrefcolor}{rgb}{0.8,0,0}
\definecolor{purple}{RGB}{128,0,128}
\definecolor{ultramarine}{RGB}{63, 0, 255}
\definecolor{medblue}{RGB}{0, 0, 100}
\definecolor{googleblue}{RGB}{34, 0, 204}
\definecolor{panblue}{RGB}{0,24,150}
\definecolor{carmine}{RGB}{150, 0, 24}
\definecolor{gray}{RGB}{150, 150, 150}

\usepackage[unicode=true,pdfusetitle, bookmarks=false,bookmarksnumbered=false,
bookmarksopen=false, breaklinks=false,pdfborder={0 0 0},backref=false,
colorlinks=true,
linkcolor=carmine,
citecolor=googleblue,
urlcolor=panblue,
anchorcolor=OliveGreen,
]{hyperref}

\usepackage{paralist}

\newcommand{\figref}[1]{Fig.~\ref{#1}}

\newcommand{\avg}[1]{\left\langle #1 \right\rangle}

\newcommand{\Tr}{\text{Tr}}

\usepackage{microtype}

\begin{document}


\title{Quantum violations in the Instrumental scenario
and their relations to the Bell scenario}

\author{Thomas Van Himbeeck}
\affiliation{Laboratoire d'Information Quantique, Universit\'e Libre de Bruxelles, 1050 Bruxelles, Belgium}
\affiliation{Centre for Quantum Information \& Communication, Universit\'e Libre de Bruxelles, Belgium}

\author{Jonatan Bohr Brask}
\affiliation{Department of Applied Physics, University of Geneva, 1211 Geneva, Switzerland}

\author{Stefano Pironio}
\affiliation{Laboratoire d'Information Quantique, Universit\'e Libre de Bruxelles, 1050 Bruxelles, Belgium}

\author{Ravishankar Ramanathan}
\affiliation{Laboratoire d'Information Quantique, Universit\'e Libre de Bruxelles, 1050 Bruxelles, Belgium}

\author{Ana Bel{\'e}n Sainz}
\affiliation{Perimeter Institute for Theoretical Physics, 31 Caroline St. N, Waterloo, Ontario, Canada, N2L 2Y5}
\affiliation{International Centre for Theory of Quantum Technologies, University of Gda\'nsk, 80-308 Gda\'nsk, Poland}

\author{Elie Wolfe}
\affiliation{Perimeter Institute for Theoretical Physics, 31 Caroline St. N, Waterloo, Ontario, Canada, N2L 2Y5}

\date{\today}

\begin{abstract}
The causal structure of any experiment implies restrictions on the observable correlations between measurement outcomes, which are different for experiments exploiting classical, quantum, or post-quantum resources. In the study of Bell nonlocality, these differences have been explored in great detail for more and more involved causal structures. Here, we go in the opposite direction and identify the simplest causal structure which exhibits a separation between classical, quantum, and post-quantum correlations. It arises in the so-called Instrumental scenario, known from classical causal models. We derive inequalities for this scenario and show that they are closely related to well-known Bell inequalities, such as the Clauser-Horne-Shimony-Holt inequality, which enables us to easily identify their classical, quantum, and post-quantum bounds as well as strategies violating the first two. The relations that we uncover imply that the quantum or post-quantum advantages witnessed by the violation of our Instrumental inequalities are not fundamentally different from those witnessed by the violations of standard inequalities in the usual Bell scenario. However, non-classical tests in the Instrumental scenario require fewer input choices than their Bell scenario counterpart, which may have potential implications for device-independent protocols.
\end{abstract}

\maketitle


\section{Introduction}

Classical and quantum physics provide fundamentally different predictions about the correlation which can be observed in experiments with multiple parties. Understanding the exact nature of this difference is a central problem in the foundations of quantum physics and is also important for applications in information processing.

In any experiment, the causal structure of the setup imposes restrictions on the observable correlations. Depending on whether the experiment is modeled using classical random variables, quantum states and measurements, or post-quantum resources, these limitations may be different, leading to observable differences between classical models, quantum mechanics, and general probabilistic theories. This was first pointed out by Bell \cite{Bell1964}, who found that models which attempt to describe an experiment in terms of causal relations between classical random variables, and where the actions of one party cannot influence the local observations of separate parties, imply restrictions on the observable correlations, known as Bell inequalities. Measurements on entangled quantum states shared between the observers, on the other hand, can lead to violation of these inequalities.

This discovery sparked the study of Bell nonlocality which by now is an active field of research and a cornerstone of quantum theory \cite{Brunner2014}. Bell's original setting involves two non-communicating parties each selecting a measurement to perform, and each obtaining a measurement outcome. Later studies have considered many variations of this causal structure, for example, {multiple parties,}  sequential measurements \cite{Popescu1995}, multiple sources \cite{Branciard2010,Fritz2012}, and communication between the parties \cite{Toner2003,Brask2017}{, as well as broader generalisations of causal modeling to quantum systems, starting from key concepts in classical causality such as Reichenbach's principle \cite{Allen2017} or the causal Markov condition \cite{Costa2016}}. In general, the causal structures which have been studied are more complicated than Bell's original setting. Here, we go in the opposite direction and identify the simplest causal structure that exhibits a separation between classical, quantum, and post-quantum correlations.

\begin{figure}[t]
\centering
\includegraphics[width=\columnwidth]{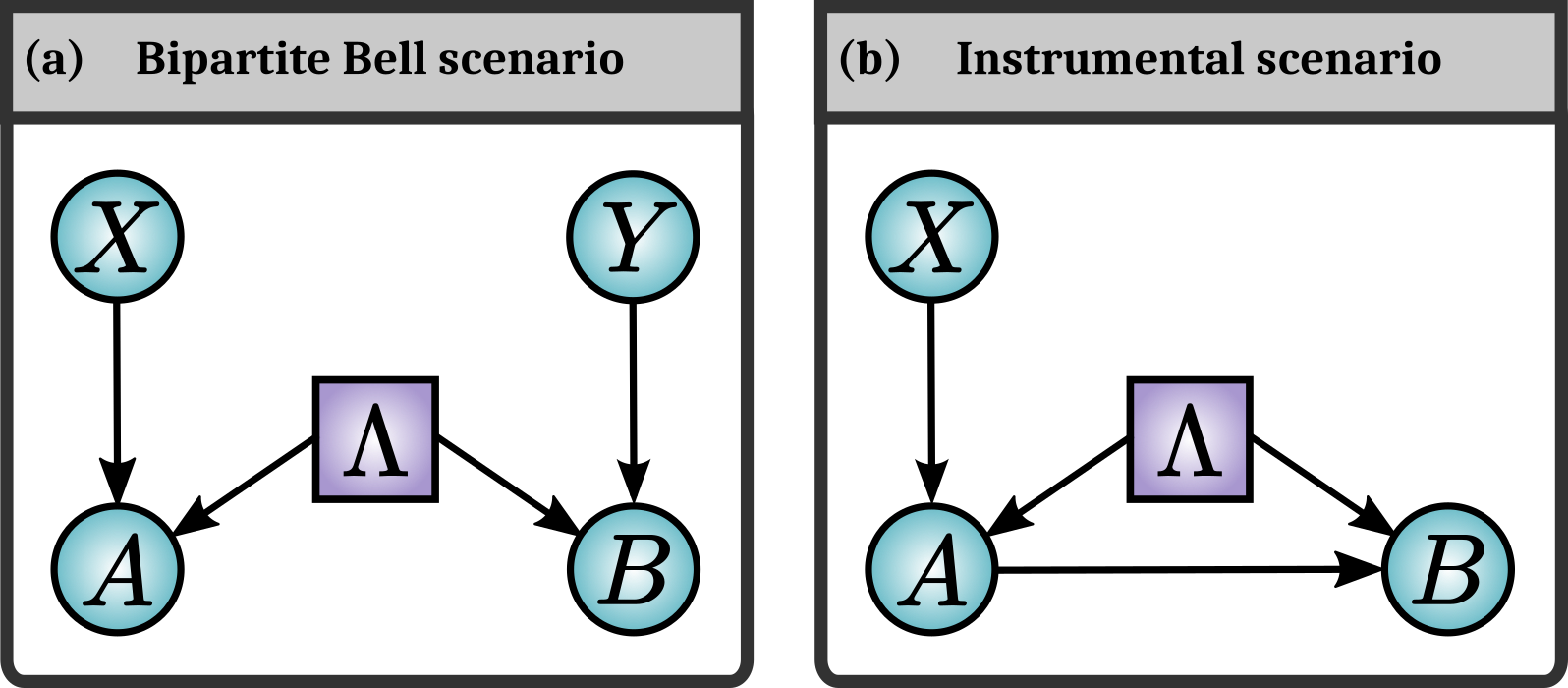}
\caption{DAGs for the standard Bell scenario, and the instrumental scenario. Circles and squares denote observable and unobservable variables respectively. Arrows denote causal influence. \textbf{(a)} Bipartite Bell scenario. Each party has an input ($X$ and $Y)$ and an output ($A$ and $B$), which are observable. An unobservable shared source $\Lambda$ may influence the outputs. \textbf{(b)} Instrumental scenario. The second party has no input, but the output of the first party is communicated to the second.}
  \label{fig.DAGs}
\end{figure}

To be a little more specific about what we mean by `simple', let us first note how a causal structure can be represented. In a given experiment, there is a number of observable variables. For instance, on a measurement apparatus, one variable (the `input') may correspond to the setting of a knob determining the measurement to be made, and another variable (the `output') to the measurement outcome. In addition, there may be hidden variables, which are not observed, but which mediate correlations between the observable variables. For instance, setting the knob on an apparatus in one way may determine the reading of a distant apparatus through an unobserved electromagnetic field.  We can represent the causal relationships between all these variables on a directed acyclic graph (DAG), where the nodes correspond to variables, and the edges between them signify causal influence. For classical models, all the variables are classical random variables. For quantum or post-quantum models, the hidden variables may be replaced by quantum states or more generally by the resources of generalized probabilistic theories (GPT) \cite{Barrett2007}, such as Popescu-Rohrlich boxes \cite{Popescu1994}. {Note that in all cases, there are two types of nodes corresponding to observed and unobserved (i.e., hidden) variables.} \figref{fig.DAGs}(a) shows the DAG for the standard, bipartite Bell scenario. {We consider a causal structure to be simpler if it has fewer nodes and edges}.

The causal structure, represented by the DAG, constrains the possible correlations between the observed variables, which further depends on whether the hidden variables are classical, quantum or GPT. {Observed variables with no parent nodes in the DAG can be thought of as inputs which are under the experimenter's control. The interesting correlations, characterizing the behaviour of a device, are therefore the conditional probabilities of the the remaining variables given these inputs}. For instance, in the Bell DAG of \figref{fig.DAGs}(a) the correlations between the observed random variables $A,B,X,Y$ are characterized by the conditional probabilities {$p=\{p(ab|xy)\}$}, which in the classical case take the form:
\begin{equation}\label{eq:class}
p\in \mathcal{C}_{\text{Bell}}\quad \textit{iff}\quad
p(ab|xy) = \sum_\lambda p(\lambda) p(a|x\lambda) p(b|y\lambda)\, .
\end{equation}
This is simply the usual Bell locality condition, and it leads to linear constraints on $p(ab|xy)$, which are Bell inequalities.

 In the quantum case,
\begin{equation}\label{eq:q}
p\in\mathcal{Q}_{\text{Bell}}\quad \textit{iff}\quad
p(ab|xy) = \text{tr}\left(\rho\, E_{a|x}\otimes F_{b|y}\right)\, ,
\end{equation}
where $\rho$ denotes a quantum state produced by $\Lambda$ and distributed to the quantum devices in $A$ and $B$; for each $x$, $\{E_{a|x}\}_a$ is a POVM defining a valid measurement with outcomes $a$; and for each $y$, $\{F_{b|y}\}_b$ is a POVM defining a valid measurement with outcomes $b$.

For the GPT case,
\begin{IEEEeqnarray}{rL}
	p\in\mathcal{G}_{\text{Bell}}\quad \textit{iff}\quad &p(ab|xy) =  (e_{a|x}|(e_{b|y}|\circ|\Psi)\,, \IEEEnonumber\\
		&{\textit{for some GPT,}} \label{eq:5}
\end{IEEEeqnarray}
where, using the notation of \cite{chiribella_probabilistic_2010}, $|\Psi)$ denotes a GPT generalization of the quantum state $\rho$ in (\ref{eq:q}), and $(e_{a|x}|$, $(e_{b|y}|$ GPT generalizations of the quantum measurement operators $E_{a|x}$, $E_{b|y}$. An example of a GPT beyond quantum theory is the one known as boxworld \cite{Barrett2007,Gross2010}, and the set of such GPT correlations for the Bell scenario coincides with the set of no-signalling correlations.

These definitions can be generalized to arbitrary DAGs beyond the Bell scenario. The classical case has been studied extensively in the classical causality literature \cite{Pearlbook}. Definitions of quantum and GPT correlations for arbitary DAGs were introduced by Henson, Lal, and Pusey (HLP) \cite{Henson2014}. We will not present the HLP formalism in detail, as we will not need it, and refer the interested reader to their paper. It suffices to say that when thinking of a set of correlations, be it classical, quantum, or any other GPT, we can think of it as arising from `measurements' being performed on a `state', where the measurements and state are dubbed classical, quantum or GPT.
Since classical correlations are a particular case of quantum correlations, and since quantum mechanics a particular case of a GPT, it follows that the sets of correlations associated with the various generalization of a causal structure form a hierarchy 
\begin{align}
\mathcal{C}_{\text{DAG}} \subseteq \mathcal{Q}_{\text{DAG}} \subseteq \mathcal{G}_{\text{DAG}} \, ,
\end{align}
{where $\mathcal{G}_{\text{DAG}}$ is the set of all correlations compatible with any GPT, not just those compatible with a specific GPT.}

While the classical, quantum, and GPT sets are strictly distinct in the Bell scenario, this is not always the case for an arbitrary DAG. HLP have introduced a necessary condition for these three sets to be distinct. Given a DAG, one can thus evaluate the HLP condition. If this condition is not satisfied, then the sets of classical, quantum, and GPT correlations are equal, i.e., the causal structure represented by the DAG is uninteresting as it does not lead to observable differences between these theories. If the HLP condition is satisfied, then one cannot conclude anything yet: classical, quantum, and GPT models might lead to observable differences, or might not  -- some further analysis is required.

In their paper, HLP have applied their criterion to all possible DAGs with 7 nodes or less \cite{Henson2014}, identifying all DAGs that \textit{possibly} admit a separation between classical, quantum, and post-quantum correlations. They have found a single DAG that is simpler than the Bell DAG, where `simple' means that it involves fewer nodes and edges. This DAG is represented in \figref{fig.DAGs}(b). It has been studied previously in the classical causality literature, where it is known as the `Instrumental DAG' \cite{Pearl1995,Bonet2001}, a nomenclature we will follow. We show here that the Instrumental scenario does indeed provide a separation between the sets of classical, quantum, and GPT correlations (we note that quantum violations for this DAG were also found independently in \cite{chaves_quantum_2017}). We derive an inequality which must hold for classical correlations and relate it to the well known Clauser-Horne-Shimony-Holt (CHSH) Bell inequality \cite{chsh} for the scenario of \figref{fig.DAGs}(a). In so doing, we identify its maximal quantum and GPT violations. We start by describing the instrumental scenario in more detail and relating it to the Bell scenario.

\section{The instrumental scenario and its relation to the Bell scenario}\label{se:3}


{Imagine possessing some quantum implementation of the Bell scenario: Alice and Bob share a bipartite quantum state (replacing the shared variable $\Lambda$), each accept a classical input determining their measurement setting (corresponding to variables $X$ and $Y$ respectively), and each produce a classical output (corresponding to $A$ and $B$). In the Bell scenario, there can be no communication between the Alice's and Bob's labs, so no causal influences between Alice's variables and Bob's. However, we can consider taking the same implementation and using it in the instrumental scenario, where $B$ is allowed to depend on $A$, by modifying it as follows: Instead of letting Bob choose his setting $y$ freely, copy Alice's output $a$ and wire the copy into Bob's input. This creates a new device, characterized by the conditional probabilities $p(ab|x)$, since $y$ is no longer freely chosen. Of course, $p(ab|x)=p(ab|x,y=a)$ so we could describe the probabilities $p(ab|x)$ that would characterize the new device without ever needing to actually perform the hypothetical modification, so long as we have a priori knowledge of $p(ab|xy)$.}

{Thus, we see that correlations in the quantum Instrumental scenario can be obtained from correlations in the quantum Bell scenario by exploiting that communication from $A$ to $B$ becomes possible. Furthermore, the same is true for the classical and GPT variants of the scenarios, as we can simply take $\Lambda$ to be a source of either classical or GPT states.} This leads us to the following fundamental statement
\begin{equation}\label{eq:IntsrBellRelationship}
p\in\mathcal{T}_{\text{Instr}} \textit{ iff }
p(ab|x) =p'(ab|x,y{=}a)
\text{ where }p'\in\mathcal{T}_{\text{Bell}}
\end{equation}
where $\mathcal{T}$ is a placeholder for a correlation set, such as classical $\mathcal{C}$, quantum $\mathcal{Q}$, or GPT $\mathcal{G}$.

In this sense, correlations in the Instrumental scenario {can be seen as} postselections of Bell-scenario correlations: An experimenter might perform many runs of the Bell scenario experiment, but then postselect to examine only those experimental runs when ${y{=}a}$. This postselected data will exhibit Instrumental scenario statistics, and moreover, \textit{every} $\mathcal{T}_{\text{Instr}}$ correlation can arise via this sort of postselection on Bell scenario statistics\footnote{The recognition that every $\mathcal{T}_{\text{Instr}}$ correlation must be compatible with $\mathcal{T}_{\text{Bell}}$ is an example of a device-independent causal inference technique which we call \textit{Interruption}. Relating pairs of DAGs in this manner is explored more generally in a forthcoming manuscript, currently in preparation~\cite{wolfe2018gptinference}.}.

To make this relationship between DAGs more concrete, note that compatibility with the Instrumental scenario is defined nearly identically to compatibility with the Bell scenario, except all references to the variable $Y$ get overwritten with references  to the variable $A$:
\begin{IEEEeqnarray}{l?r?l}
	p\in\mathcal{C}_{\text{Instr}} &\textit{iff}
		& p(ab|x) = {\textstyle \sum_\lambda} p(\lambda) p(a|x\lambda) p(b|a\lambda)\, ; \IEEEeqnarraynumspace \\
	p\in\mathcal{Q}_{\text{Instr}} &\textit{iff}
		& p(ab|x) = \text{tr}\left(\rho\, E_{a|x}\otimes F_{b|a}\right)\, ;\\
	p\in\mathcal{G}_{\text{Instr}} &\textit{iff}
		& p(ab|x) =  (e_{a|x}|(e_{b|a}|\circ |\Psi)\,,\IEEEnonumber\\
		&&{\textit{for some GPT}}\,.
\end{IEEEeqnarray}


Even though the Quantum- and GPT-Bell-Scenario sets are in general distinct from the Classical-Bell-Scenario set, it may be that their postselections (defining the corresponding Instrumental-Scenario correlation sets) all coincide. {Indeed, it is well known that postselected classical data may reproduce postselected genuine quantum, nonlocal data.} This effect is at the basis, for instance, of the infamous detection loophole in Bell experiments \cite{Brunner2014}. It is thus not obvious a priori that the Instrumental scenario should admit a separation between classical, quantum and GPT correlations; it might be another example of an uninteresting DAG, which simply happens not to be identified by the HLP criterion.

\begin{figure}[t]
\centering
\includegraphics[scale=0.8]
{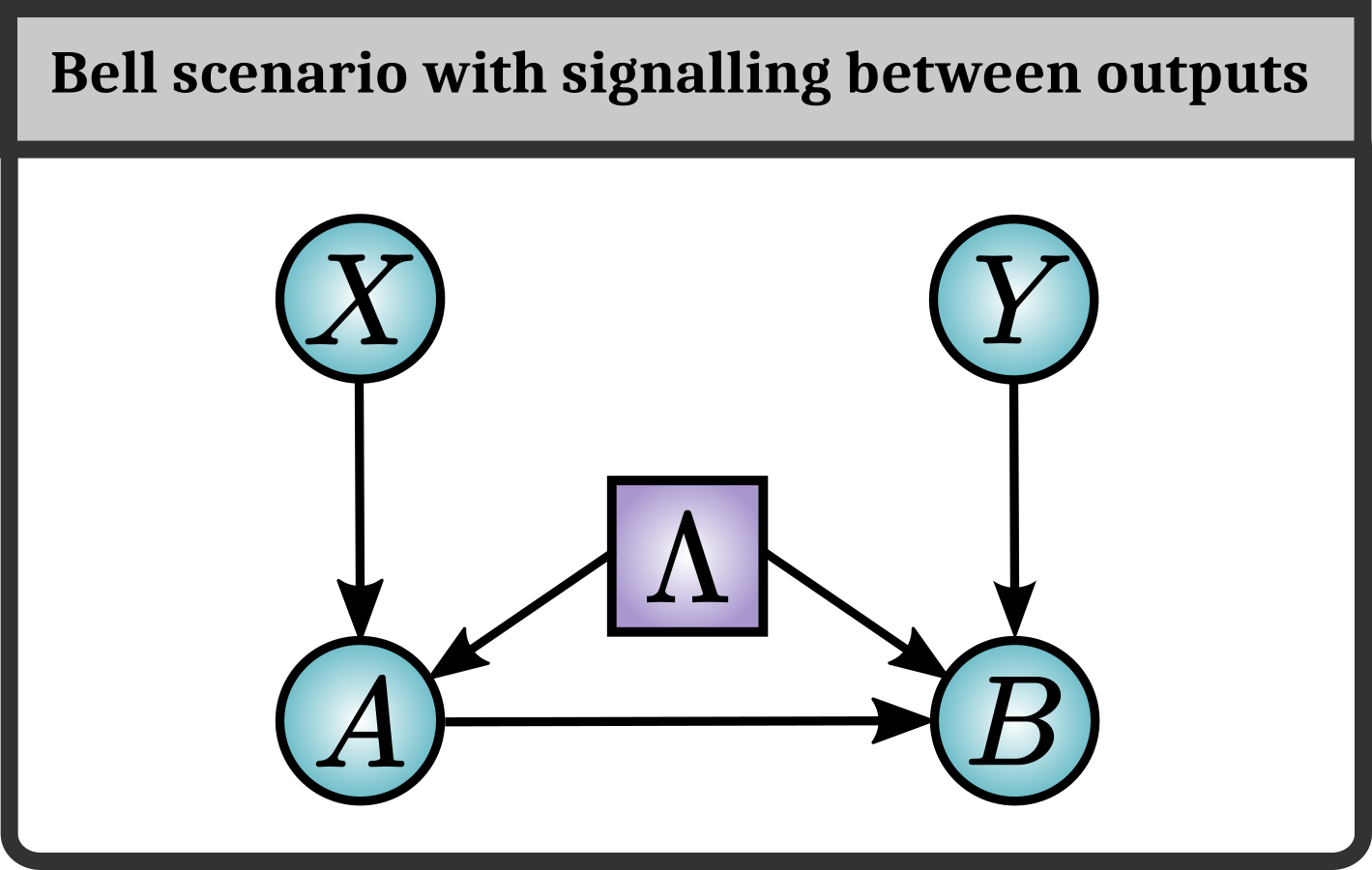}
\caption{The Signalling-Between-Outputs scenario considered in Refs.~\cite{Brask2017,Chaves2015}. This scenario relaxes the no-communication assumption between the variables $A$ and $B$, thus modifying the usual Bell scenario depicted in \figref{fig.DAGs}(a) by the addition of an edge $A\to B$.}
  \label{fig.DAG_sbo}
\end{figure}

The Instrumental scenario can also be understood as a Bell scenario with relaxed causality constraints. Such relaxations of the Bell scenario have been considered previously. For instance, the no-communication assumption between the variables $A$ and $B$ in the Bell scenario has been relaxed in Refs.~\cite{Brask2017,Chaves2015}, leading to the Signalling-Between-Outputs scenario represented in \figref{fig.DAG_sbo}. Other works have considered modified Bell scenarios  wherein one no longer assumes that the measurement inputs $X$ and $Y$ could have been set-up freely \cite{barrett_how_2011,putz_arbitrarily_2014}. The Instrumental scenario represents simultaneous relaxation of both the measurement-freedom and no-communication assumptions: not only may the outcome of the measurement performed at $B$ depend directly on the distant outcome $A$, but furthermore the measurement setting $Y$ is not chosen freely but is instead fixed entirely by the value $A$ output by the distant measurement device.

We remark that testing for membership in $\mathcal{T}_{\text{Instr}}$, by admitting an extension to $\mathcal{T}_{\text{Bell}}$ per Eq.~\eqref{eq:IntsrBellRelationship}, generalizes to any well defined correlation set in the Bell scenario{, not just $\mathcal{C}$, $\mathcal{Q}$, or GPT correlations}. For example, one might consider relaxations of the quantum set corresponding to levels of the Navascu{\'e}s-Pironio-Ac{\'{\i}}n (NPA) hierarchy \cite{NPA,NPAprl}, including the set known as Almost Quantum Correlations~\cite{AQC}, or tests for compatibility with a restricted Hilbert space dimension~\cite{Vertesi2014NPA,Vertesi2015NPA,Sikora2016HSD}. All those correlation membership tests for the Bell scenario can be applied to the Instrumental scenario by simply introducing existential quantifiers: Does an extension of $p(ab|x)\equiv p(ab|x,y{=}a)$ exist to $p(ab|x,y{\neq}a)$ such that the relevant condition for membership in $\mathcal{T}_{\text{Bell}}$ is satisfied? This modification is especially easy for testing NPA-level compatibility, as those semidefinite tests already natively support {input data where not all the probabilities are specified}.

\section{Geometry of the Instrumental-Scenario correlations}

Before attempting to find a gap between classical and quantum correlations in the instrumental scenario, let us take a general geometric perspective to enhance our understanding. Every correlation in the Bell scenario can be thought of as a high-dimensional vector, ${d=|A||B||X||Y|}$, where the coordinates are given by the many different probabilities $p(ab|xy)$. Every correlation in the Instrumental scenario can be thought of as a somewhat lower-dimensional vector, ${d=|A||B||X|}$, where the coordinates are given by the probabilities $p(ab|x)$. The set of all correlations in $\mathcal{T}_{\text{Instr}}$ are formed by axial \textit{projection} of those coordinates $p(ab|x,y{\neq}a)$ of $\mathcal{T}_{\text{Bell}}$.

In both the Bell and Instrumental scenarios, all sets of correlations are convex. The sets $\mathcal{G}_{\text{Bell}}$ and $\mathcal{C}_{\text{Bell}}$ are the no-signalling polytope and the local polytope respectively, whereas the set $\mathcal{Q}_{\text{Bell}}$ is a convex set but not a polytope~\cite{Brunner2014,Avis2007,BancalDIApproach,Scarani2012device}.
The projections of polytopes are also polytopes, so we know that $\mathcal{G}_{\text{Instr}}$ and $\mathcal{C}_{\text{Instr}}$ will also form polytopes. To obtain the Instrumental scenario polytopes from the Bell scenario polytopes, we can use Fourier-Motzkin elimination or any other polytope projection technique~\cite{DantzigEaves,BalasProjectionCone,jones2004equality,Shapot2012,Bastrakov2015}.
Alternatively, we can directly compute $\mathcal{C}_{\text{Instr}}$ by taking the convex hull of all deterministic strategies in the Instrumental scenario. We have performed these operations for small cardinalities of the observed variables $X,A,B$ using the polytope software PORTA \cite{porta}.

In the simplest case where $X,A,B\in\{0,1\}$ are all binary, we find that $\mathcal{C}_{\text{Instr}}=\mathcal{G}_{\text{Instr}}$ and that these sets are fully characterized by the trivial normalization
\begin{equation}
	\sum_{ab}p(ab|x)=1
\end{equation}
and positivity
\begin{equation}
	p(ab|x)\geq 0
\end{equation}
conditions, together with the additional set of constraints
\begin{equation}
	\label{eq.instrumental.explicit}
	{ p(a0|x)+p(a1|x') \leq 1\quad \text{for all } a,x\neq x'\,, }
\end{equation}
which can be expressed compactly as
\begin{equation}
	\label{eq.instrumental}
	\underset{a}{\max} \sum_b \underset{x}{\max} \,\, p(ab|x) \leq 1\,.
\end{equation}

As $\mathcal{C}_{\text{Instr}}\subseteq\mathcal{Q}_{\text{Instr}} \subseteq\mathcal{G}_{\text{Instr}}$, the above constraints also fully characterize the quantum set $\mathcal{Q}_{\text{Instr}}$.

Since the normalization condition is the only generic equality constraint satisfied by correlations in the Instrumental scenario, the sets $\mathcal{C}_{\text{Instr}}$, $\mathcal{Q}_{\text{Instr}}$, and $\mathcal{G}_{\text{Instr}}$ are full-dimensional in the space of normalized probability distributions. This should be contrasted with the Bell scenario where  $\mathcal{C}_{\text{Bell}}$, $\mathcal{Q}_{\text{Bell}}$, and $\mathcal{G}_{\text{Bell}}$ are not full-dimensional in the space of normalized probability distributions, since they also satisfy the no-signaling equality constraints, expressing that the marginal distribution of $b$ cannot depend on $x$ and the marginal distribution of $a$ on $y$.
This full-dimensional property of the Instrumental scenario is not limited to the ${|X|{=}|A|{=}|B|{=}2}$ case, but is valid for any cardinalities of the inputs and outputs. Indeed, a method to determine the complete set of equality constraints satisfied by classical correlations compatible with an arbitrary DAG has been given in \cite{evans2015equalityconstraints}. {One can verify that applying} it to the Instrumental DAG yields no other equalities than the normalization conditions in the classical case -- and hence in the quantum and GPT case as well, since they contain classical correlations as a subset.

Even though the Instrumental scenario does not contain no-signaling constraints -- indeed the input $b$ can depend on $x$ through $a$ -- some remnant of the no-signaling conditions are preserved when projecting the Bell scenario to the Instrumental scenario, as expressed by the inequalities (\ref{eq.instrumental.explicit}), which can be interpreted as limiting the magnitude by which $b$ can depend on $x$ when $a$ is kept constant.

We can understand that such inequalities cannot be violated by a GPT from the {fact that $\mathcal{G}_{\text{Instr}}$ is the projection of the no-signalling polytope}. As an example, let us derive one of the inequalities (\ref{eq.instrumental.explicit}) from the following two positivity inequalities for the Bell scenario\footnote{{We add subscripts indicating the variables for the conditional distributions, when there may be risk of confusion. E.g.~$p(ab|xy)$ and $p_{AB|XY}(ab|xy)$ are the same object.}} : $p_{AB|XY}(11|10)\geq 0$ and $p_{AB|XY}(10|00)\geq 0$. Summing those two inequalities together and then using no-signalling to express the probabilities as linear combinations of those where Alice's output matches Bob's input, i.e., $p_{AB|XY}(11|10)\to p_{B|Y}(1|0)-p_{AB|XY}(01|10)$ and $p_{AB|XY}(10|00)\to p_{B|Y}(0|0)-p_{AB|XY}(00|00)$, we obtain
\begin{multline}
	p_{B|Y}(0|0)+p_{B|Y}(1|0) \\-p_{AB|XY}(00|00) - p_{AB|XY}(01|10)\geq 0\,,
\end{multline}
or, equivalently,
\begin{align}
	p_{AB|XY}(00|00) + p_{AB|XY}(01|10)\leq 1\,.
\end{align}
Having eliminated the non-Instrumental probabilities $p(ab|x,y\neq a)$, the final inequality (as translated for the Instrumental scenario) reads
\begin{align}\label{eq:instrumentalnoviolation}
	p_{AB|X}(00|0) + p_{AB|X}(01|1)\leq 1\,,
\end{align}
This proves that Eq.~\eqref{eq:instrumentalnoviolation} -- an instance of (\ref{eq.instrumental.explicit}) -- is an Instrumental scenario inequality{, which cannot be violated by a GPT.}

Expressed in the general form (\ref{eq.instrumental}), these inequalities are valid for $\mathcal{C}_{\text{Instr}}$, $\mathcal{Q}_{\text{Instr}}$, and $\mathcal{G}_{\text{Instr}}$ for arbitrary number of inputs and outputs $|X|$, $|A|$, $|B|$. They were originally derived by Pearl~\cite{Pearl1995} for the classical Instrumental scenario and have come to be known as \textit{the} instrumental inequalities. Henson, Lal, and Pusey then showed that Pearl's instrumental inequalities (\ref{eq.instrumental}) are satisfied by all GPTs for
arbitrary inputs and outputs \cite{Henson2014}.

To summarize, we have found that in the case ${|X|{=}|A|{=}|B|{=}2}$ the instrumental inequalities (\ref{eq.instrumental}) are the unique facets, besides the trivial positivity facets, of the GPT polytope. We have verified that this is also the case for $|X|{=}2$ and ${|A|{=}|B|\leq 4}$. We also know that the instrumental inequalities are satisfied by the GPT polytope for arbitrary number of inputs and outputs, but we leave it as on open question whether they are the unique non-trivial facets {when $|X|=2$}.

In the simplest possible Bell scenario where ${X{,}Y\!{,}A{,}B\in\{0,1\}}$ well-known bounds on the violation of the CHSH inequality imply that ${\mathcal{C}_{\text{Bell}}\subsetneq\mathcal{Q}_{\text{Bell}}\subsetneq \mathcal{G}_{\text{Bell}}}$. We have found, however, that ${\mathcal{C}_{\text{Instr}}=\mathcal{Q}_{\text{Instr}} =\mathcal{G}_{\text{Instr}}}$ for the corresponding Instrumental sets, i.e., all non-classical features of Bell correlations are washed out when post-selecting them to obtain the Instrumental correlations of the ${X{,}A{,}B\in\{0,1\}}$ set-up. Though, we have established this fact by fully characterizing the Instrumental polytopes using the software PORTA, it is also instructive to see more explicitly how all non-local correlations of the ${X{,}Y\!{,}A{,}B\in\{0,1\}}$ Bell scenario admit a classical explanation when projected to the Instrumental scenario. Consider for instance the Popescu-Rorhlich (PR) correlations
\begin{equation}
	p(ab|xy) = \left\{\begin{array}{lr} 1/2& \text{if } b=a+xy {\mod 2}\\
	0&\text{otherwise}\,, \end{array}\right.
\end{equation}
which reaches the {maximal} algebraic value of $4$ of the CHSH expression {(given in \eqref{eq.CHSH} below)}.
Post-selecting the case where $p(ab|x,y=a)$, we get the following Instrumental scenario correlations
\begin{equation}
	p(ab|x) = \left\{\begin{array}{lr} 1/2& \text{if } b=a(1+x) {\mod 2}\\
	0&\text{otherwise\,.} \end{array}\right.
\end{equation}
In other words, $p(a|x)=1/2$ and ${b=a}$ if ${x{=}0}$, while ${b=0}$ if ${x{=}1}$. But now it is easy to see how these correlations can be simulated classically. Consider a binary hidden variable $\lambda\in\{0,1\}$ that is unbiased, i.e., ${p(\lambda{=}0)=p(\lambda{=}1)=1/2}$, and define $a:=\lambda+x$, ${b:=\lambda\, a}$ {(with addition modulo 2)}. We obviously have that $p(a|x)=1/2$ and ${b=\lambda^2{=}\lambda{=}a}$ if ${x{=}0}$, while ${b=\lambda(\lambda+1){=}0}$ if ${x{=}1}$, as required.

Since any GPT correlations in the Bell scenario can be written as a mixture of classical correlations and PR correlations, any GPT correlations {in the Instrumental scenario} can be written as a mixture of classical correlations and post-selected PR correlations. But since we have just seen that the later ones are classical, this establishes that $\mathcal{G}_{\text{Instr}}=\mathcal{C}_{\text{Instr}}$. More generally, it was shown in \cite{Brask2017,Chaves2015} by a similar argument that classical models can reproduce any GPT correlations in the Signalling-Between-Outputs scenario whenever ${|X|{=}|Y|{=}|A|{=}2}$ and $|B|$ is arbitrary. These results translate to our case since the Instrumental DAG is a subgraph of the Signalling-Between-Outputs DAG in which the node $Y$ is dropped. They imply that there cannot be any separation between classical, quantum and GPT correlations in the Instrumental scenario whenever ${|X|{=}|A|{=}2}$ and $|B|$ is arbitrary.

\section{A classical Instrumental scenario inequality which admits quantum violation}\label{se:4}
{The case $X\in\{0,1,2\}$, $A,B\in\{0,1\}$ is more interesting, as Bonet \cite{Bonet2001} found (also using PORTA) that the facets of the classical polytope $\mathcal{C}_{\text{Instr}}$ comprise, in addition to Pearl's instrumental inequalities, a new family of inequalities, one representative of which is}
\begin{equation}
	\label{eq.newineq}I_{\text{Bonet}}:=
	p(a{=}b|0)+p(b{=}0|1)+p(a{=}0,b{=}1|2) \leq 2 .
\end{equation}

This inequality admits quantum violation, and moreover also provides quantum/GPT separation, as we now show. A quantum strategy violating (\ref{eq.newineq}) is as follows. Let the source $\Lambda$ distribute the two-qubit maximally entangled state $|\phi_+\rangle=(|00\rangle+|11\rangle)/\sqrt{2}$ and let {Alice} perform the measurements $\sigma_x,\sigma_z,-(\sigma_x+\sigma_z)/\sqrt{2}$ when she receives the input $X=0,1,2$, respectively, while Bob measures $(\sigma_x+\sigma_z)/\sqrt{2},(\sigma_x-\sigma_z)/\sqrt{2}$ when he receives $a=0,1$. A straightforward computation gives $I_{\text{Bonet}}=(3+\sqrt{2})/2\simeq 2.207>2$.

An example of GPT correlations violating Bonet's inequality is given by
\begin{equation}
p(ab|x) = \left\{\begin{array}{ll} 1/2& \text{if } b=a+f(x,a) {\mod 2}\\
0&\text{otherwise\,,} \end{array}\right.
\end{equation}
where {$f(0,a)=0$, $f(1,a)=a$,} and $f(2,a)=a+1 {\mod 2}$. Inserting these probabilities in (\ref{eq.newineq}) yields $I_{\text{Bonet}}=5/2>2$. It can be verified that these correlations are GPT valid as they satisfy Pearl's instrumental inequalities, which are the unique (non-trivial) facets of the GPT polytope in the $|X|=3$, $|A|=|B|=2$ case. Alternatively, they can be seen as postselection of the GPT (i.e., no-signalling) Bell correlations $p(ab|xy)$ defined by $p(ab|xy)=1/2$ if $b=a+f(x,y) {\mod 2}$ and $0$ otherwise.

\section{Relation to the CHSH inequality and dummy inputs}\label{se:6}

The fact that post-selections of the ${|X|{=}|Y|{=}|A|{=}|B|{=}2}$ Bell scenario, where non-locality is entirely detected by the violation of the CHSH inequality, do not lead to non-classical Instrumental correlations might suggest, naively, that violation of Bonet's inequality (\ref{eq.newineq}) uncover a stronger form of non-locality, requiring violating beyond the CHSH inequality. We show that this is not the case by relating Bonet's inequality to the CHSH inequality. That such a link must exist also follows directly from the fact that all (non-trivial) facets of the ${|X|{=}3}$, ${|Y|{=}|A|{=}|B|{=}2}$ classical Bell polytopes are liftings of the CHSH inequality \cite{pironio_all_2014}.

Although we found inequality \eqref{eq.newineq} by taking the convex hull of deterministic strategies and without regard to the relationship between the Bell and Instrumental scenarios, it is enlightening to {retrospectively} explain $I_{\text{Bonet}}$ as a projection of the classical Bell scenario polytope.

Let us rewrite the expression $I_{\text{Bonet}}$ per \eqref{eq.newineq} in terms of $p(ab|xy)$; that is, let us interpret the facet of the classical Instrumental polytope as a valid inequality for the Bell polytope. This operation is a trivial lifting of the inequality via the mapping $p(ab|x)\to p(ab|x,y{=}a)$. We find that
\begin{equation}
	\begin{split}
		\textrm{Lifting}\left[ I_{\text{Bonet}} \right] = \, & p(00|00) + p(11|01) + p(00|10) \\
		& + p(10|11) + p(01|20)\,.
	\end{split}
\end{equation}
Using the normalization and no-signalling constraints satisfied by the Bell scenario probabilities $p_{AB|XY}$, we can rewrite this last expression as
\begin{equation}
	\label{eq.liftedBonet}
	\textrm{Lifting}\left[ I_{\text{Bonet}} \right]  = \frac{1}{4}\avg{CHSH} -  p(11|20)+\frac{3}{2} ,
\end{equation}
where
\begin{equation}
\label{eq.CHSH}
\avg{CHSH}\equiv \avg{A_0B_0}+\avg{A_0B_1}+\avg{A_1B_0}-\avg{A_1B_1} ,
\end{equation}
with $\avg{A_xB_y}$ the expectation value of $(-1)^{A+B}$ given that $X$ and $Y$ take values $x$ and $y$ respectively. From {\eqref{eq.liftedBonet}} it becomes immediately clear that the classical, quantum, and GPT bounds of $I_{\text{Bonet}}$ are
\begin{align}
I_{\text{Bonet}} \leq \begin{cases}
        2 & \text{Classical}\\
        (3+\sqrt{2})/2 & \text{Quantum}\\
        5/2 & \text{GPT}
    \end{cases}\,,
\end{align}
as this follows from
\begin{align}
\avg{CHSH} \leq \begin{cases}
        2 & \text{Classical}\\
        2\sqrt{2} & \text{Quantum}\\
        4 & \text{GPT}
    \end{cases}\,,
\end{align}
as well as the fact that $-p(11|20)\leq 0$ in all physical theories.

A perhaps surprising consequence of the {retrospective} interpretation of $I_{\text{Bonet}}$ is that \textit{any} nonclassical correlation in the CHSH Bell scenario can be used as a resource to generate nonclassical correlations in the Instrumental scenario, despite the fact that the Instrumental scenario has coinciding GPT and classical polytopes for ${|X|{=}|A|{=}2}$. The trick which allows us to map arbitrary non-classical No-Signalling correlations in the Bell scenario where ${|X|{=}|Y|{=}|A|{=}|B|{=}2}$ into \textit{non}-classical correlations in the Instrumental scenario is as follows: Starting from a $p_{AB|XY}$ in the standard CHSH scenario where $x\in\{0,1\}$, trivially map it to $p'_{AB|XY}$ in an extended scenario where $x\in\{0,1,2\}$ by setting $p'(ab|xy)=p(ab|xy)$ when $x{=}0,1$ and $p'(ab|x{=}2,y)=\delta_{a,0}p(b|y)$ when $x{=}2$. That is, in the case $x=2$, the output $a$ is deterministically equal to 0. Then we have $p'(11|20)=0$ and thus we may substitute
\begin{equation}
	\frac{1}{4}\avg{CHSH}_{p'}+p'(11|20)\to \frac{1}{4}\avg{CHSH}_{p}
\end{equation}
to recast $\textrm{Lifting}\left[ I_{\text{Bonet}} \right]_{p'}$ for $|X|=3$ as an explicit function of $p$ for $|X|=2$, with the trivial intermediate map $p\rightarrow p'$ taken for granted:
\begin{equation}
	\label{eq.newineqchsh2}
	\textrm{Lifting}\left[ I_{\text{Bonet}} \right]_{p'}  = \frac{1}{4}\avg{CHSH}_p  +\frac{3}{2} ,
\end{equation}
In particular, this trivial map allows us to relate the extent of the violation of $I_{\text{Bonet}}\leq 2$ in the Instrumental scenario \textit{entirely} as a function of the extent of the violation of $\avg{CHSH}\leq 2$ in the Bell scenario. A direct consequence is that any non-classical correlations in the $|X|=|A|=|Y|=|B|=2$ Bell scenario, which necessarily violate the CHSH inequality,  give rise to non-classical correlations violating Bonet's inequality in the Instrumental scenario.

Another way to express this connection is as follows. Writing $p'$ in term of $p$, we can rewrite the relation (\ref{eq.newineqchsh2}) explicitly as the identity
\begin{align}
\label{eq.newineqchsh3}
\frac{1}{4}\avg{CHSH}_p+\frac{3}{2} &=p(00|00) + p(11|01)+ p(00|10) \nonumber\\& \, + p(10|11) + p_{B|Y}(1|0)\,.
\end{align}
Thus, instead of testing CHSH in the regular way, which involves estimating the correlations for 4 choices of input pairs $(x,y)\in\{(0,0),(0,1),(1,0),(1,1)\}$, one can alternatively test it using 3 choices of an input $z$. $(i)$ If $z{=}0,1$, one uses $x{=}z$ on Alice's side and uses Alice's outputs as an input for Bob. This allows to evaluate the first four terms on the {right-hand} side of (\ref{eq.newineqchsh3}). $(ii)$ If $z{=}2$, one uses $y{=}0$ as an input for Bob and registers his output (without testing Alice). This allows to evaluate the last term of (\ref{eq.newineqchsh3}).

\section{General mapping between Bell and Instrumental inequalities in the case $|A|=|Y|=2$}
\label{se:tilted}

The results of the last section show that at least in the specific input-output configuration we considered, the Instrumental scenario is essentially equivalent to the Bell scenario for the purpose of detecting non-classicality, i.e., correlations in the CHSH Bell scenario are non-classical if and only if they give rise to non-classical correlations in the Instrumental scenario.

However, many interesting properties of non-classical correlations do not merely reduce to testing their non-classicality, i.e., to testing their Bell inequality violation. For instance, the tilted CHSH inequalities introduced in \cite{AMP}, though weaker than the CHSH inequality for detecting non-classicality, are useful for other purposes. In a device-independent setting, they enable certifying more randomness that would be possible using standard CHSH \cite{AMP,BMP}.

We now generalize the results of the previous section and show that, starting from \textit{any} Bell inequality in the $|A|=|Y|=2$ Bell-scenario (with $|X|$ and $|B|$ arbitrary), one can build a corresponding Instrumental inequality, which can have up to $|X|+2$ inputs for Alice.

An arbitrary linear Bell functional is given by
\begin{equation}
	I_{\text{Bell}} = \sum_{a,b,x,y} \alpha_{abxy} p(ab|xy)\,.
\end{equation}
Using that $A,Y\in\{0,1\}$, we can write this expression as
\begin{IEEEeqnarray}{Rl}\label{eq:gen_ineq_Bell}
I_{\text{Bell}}&=\sum_{b,x,y} \left(\alpha_{ybxy} p(yb|xy)+\alpha_{\bar{y}bxy} p(\bar{y} b|xy)\right)\\
&=\sum_{b,x,y} \left(\alpha_{ybxy}-\alpha_{\bar{y}bxy}\right) p(yb|xy)+\sum_{b,y}\alpha_{\bar{y}by} p_{B|Y}(b|y)\IEEEnonumber
\end{IEEEeqnarray}
where $\bar y$ denotes negation of $y$, we define $\alpha_{aby}=\sum_x\alpha_{abxy}$, and we used that $p(\bar{y}b|xy)=p_{B|Y}(y|b)-p(yb|xy)$.

This expression now involves valid instrumental probabilities $p(yb|xy)$ together with the marginal probabilities $p_{B|Y}(b|y)$ which are not directly accessible in the Instrumental scenario. In order to construct an instrumental inequality, we use the fact that
\begin{equation}\label{eq.bounds}
p_{A|X}(\bar{y}|x)+p(yb|xy)\geq p_{B|Y}(b|y) \geq p(yb|xy)
\end{equation}
for any input $x$. In particular, we can introduce two new input values $\{x_0,x_1\}$ for $X$ and make the following replacement in \eqref{eq:gen_ineq_Bell}
\begin{IEEEeqnarray}{rL}
	\label{eq:ineq}
\hspace{-0.7cm}	p_{B|Y}(b|y) \rightarrow p(yb|x_y y)&\text{ if } \alpha_{\bar{y}by}> 0,\IEEEnonumber\\
	p_{B|Y}(b|y) \rightarrow p(yb|x_y y)+p_{A|X}(\bar{y}|x_y)&\text{ if } \alpha_{\bar{y}by}<0. \IEEEeqnarraynumspace
\end{IEEEeqnarray}
This results in the Bell expression
\begin{IEEEeqnarray}{rCl}
	\label{eq:gen_ineq_Instr_Bell}
\hspace{-0.7cm}	I_{\text{Instr}} &=&	\sum_{b,x,y} \left(\alpha_{ybxy}-\alpha_{\bar{y}bxy}\right) p(yb|xy)\IEEEnonumber\\
	&&+\sum_{b,y}\alpha_{\bar{y}by} p(yb|x_y y)+\sum_{b,y}\tilde\alpha_{\bar{y}by} p_{A|X}(\bar{y}|x_y),
\end{IEEEeqnarray}
where $\tilde\alpha_{\bar{y}by}=\alpha_{\bar{y}by}$ if $\alpha_{\bar{y}by}<0$ and $\tilde\alpha_{\bar{y}by}=0$ otherwise.
This expression now only involves instrumental probabilities and thus we can formally view it as the Instrumental expression
\begin{IEEEeqnarray}{rCl}
	\label{eq:gen_ineq_Instr}
	I_{\text{Instr}} &=&	\sum_{b,x,y} \left(\alpha_{ybxy}-\alpha_{\bar{y}bxy}\right) p(yb|x)\IEEEnonumber\\
	&&+\sum_{b,y}\alpha_{\bar{y}by} p(yb|x_y)+\sum_{b,y}\tilde\alpha_{\bar{y}by} p(\bar{y}|x_y)\,.\IEEEeqnarraynumspace
\end{IEEEeqnarray}
(In the following, we do not distinguish $I_{\text{Instr}}$ written in the Bell form \eqref{eq:gen_ineq_Instr_Bell} or in the Instrumental form \eqref{eq:gen_ineq_Instr}, since they are essentially equivalent).
We clearly have
\begin{IEEEeqnarray}{rL}
	\label{eq:yet_another_ineq}
	I_{\text{Instr}}[p] \leq I_{\text{Bell}}[p]
\end{IEEEeqnarray}
as follows from (\ref{eq.bounds}).

The new inequality $I_{\text{Instr}}$ has the interesting property that for any GPT $\mathcal{G}$,
\begin{IEEEeqnarray}{rL}
	\label{eq:relation_Bell_Instr}
	\max_{p\in \mathcal{G}_{\text{Bell}}} I_{\text{Bell}}[p]
	= \max_{p\in \mathcal{G}_{\text{Instr}}} I_{\text{Instr}}[p]\,.
\end{IEEEeqnarray}

This can be seen from the fact that, starting from a given resource $p(ab|xy)$ in the original Bell scenario, one can always add two dummy inputs $x_k$ for $k=0,1$ and do a simple wiring such that $p_{A|X}(a|x_k) = \delta_{a,k}$. This results in a new distribution $p'$ such that $p'(ab|x_ky) = \delta_{a,k}p(b|y)$, $p'(a|x_k)=\delta_{a,k}$ and $p'(ab|xy) = p(ab|xy)$ for the original values of $x$. Then, using this newly defined resource, we find $I_{\text{Instr}}[p']= I_{\text{Bell}}[p]$, because in the expression \eqref{eq:gen_ineq_Instr_Bell} the term $p(\bar{y}|x_y)$ disappears and $p(yb|x_y y)$ becomes equal to $p(b|y)$ so that the whole expression becomes equal to \eqref{eq:gen_ineq_Bell}.

Note that the above trick only works because we introduced an additional dummy input $x_y$ for each $p(b|y)$ appearing in \eqref{eq:gen_ineq_Bell}. If we had used instead one of the already existing inputs $x$, the inequality \eqref{eq:gen_ineq_Instr} would still hold, but could not be saturated by the same distribution that maximizes $I_{\text{Bell}}$ and we would not necessarily be able to construct a $p'$ such that $I_{\text{Instr}}[p'] = I_{\text{Bell}}[p]$.

Note also that the dummy input $x_y$ only need to be introduced if the coefficients $\alpha_{\bar{y}by}$ in front of $p(b|y)$ are non-zero in \eqref{eq:gen_ineq_Bell}.
Coming back to the CHSH inequality (and the corresponding Instrumental inequality), we notice that it satisfies $\alpha_{1b0}\neq 0$, but $\alpha_{0b1}=0$ and thus requires the introduction of a single additional input $x_0$.

As a by-product of the previous discussion, we also see that Bell inequalities in the $|X|=|A|=|Y|=|B|=2$ scenario that satisfy $\alpha_{\bar{y}by}=0$ for all $b,y$, cannot be used to test for non-locality. Indeed they correspond to instrumental inequalities with only two inputs $|X|=2$, and we have argued above that in this case there is no gap between the classical and GPT sets of correlations. Because of \eqref{eq:relation_Bell_Instr}, the absence of a gap must also hold for the original Bell-inequality.

The general mapping that we just introduced can be used to construct, e.g., instrumental versions of the tilted CHSH inequalities.

 Interestingly, {this construction of instrumental inequalities that we developed} preserves, at the point of maximal violation, all quantum properties of the original inequalities involving the inputs that are not dummy. Indeed, let $p$ be some instrumental quantum correlations that attain the maximum $I_{\text{Instr}}[p] = \max_{p\in \mathcal{Q}_{\text{Bell}}}I_{\text{Bell}}[p]$.  Such  correlations are defined by an optimum quantum strategy specified by a bipartite state $\rho$, a set of measurement operators $E_{a|x}$ for Alice and a set of measurement operators $F_{b|y}$ for Bob, such that $p(yb|xy)=\Tr\left(\rho E_{y|x}\otimes F_{b|y}\right)$. The same states and operators also define a quantum strategy $p(ab|xy)=\Tr\left(\rho E_{a|x}\otimes F_{b|y}\right)$
in the original Bell scenario, obtained by neglecting the cases $x=x_0$ and $x=x_1$. We clearly have $I_{\text{Instr}}[p]\leq I_{\text{Bell}}[p]$ because of the inequalities \eqref{eq.bounds} (with saturation possible using the deterministic assignment to dummy inputs defined above). Furthermore $I_{\text{Bell}}[p]\leq \max_{p\in \mathcal{Q}_{\text{Bell}}}I_{\text{Bell}}[p]$,
with equality only if the quantum strategy is actually optimal for the Bell scenario. Thus the Instrumental correlations $p$ can only achieve the maximum of the inequality $I_{\text{Instr}}$ if the subset of probabilities corresponding to the non-dummy inputs are optimal for the corresponding inequality in the Bell scenario. In particular, if the original Bell inequality is self-testing, i.e., up to local isometries only one quantum strategy $\{\rho,E_{a|x},F_{b|y}\}$ can violate it maximally \cite{st1}, then the Instrumental optimal strategy is necessarily of the same form, with added measurement operators for the dummy inputs.

Finally, though the mapping from the Bell to the Instrumental scenario that we just presented is restricted to the case $|A|=|Y|=2$, it may also be possible to find similar mappings in other cases. As an illustration, we show in Appendix~\ref{sec:appendix} how it is possible to map the chained Bell inequalities, corresponding to $|A|=2$, $|Y|=n$, to a (slightly general) Instrumental scenario.

\section{Discussion}\label{se:7}

The original motivation of Bell was to provide a testable criterion for whether Nature is compatible with a classical local causal description. In such an experiment, ideally one does not wish to make any assumptions about the non-existence of spurious communication channels between the parties, which could be mediated via as-yet-undiscovered physics. To rule out communication, which could explain the observed correlations, one can instead arrange to have space-like separation of the different parties' measurement events. Any communication would then need to be superluminal, in violation of special relativity. The minimal causal structure in which such an experiment can be implemented is that of \figref{fig.DAGs}(a), and the minimal scenario is that of CHSH (binary inputs and outputs for each party). Several conclusive tests imposing space-like separation have recently been realised \cite{Hensen2015,Shalm2015,Giustina2015}.

However, one of the consequences of Bell nonlocality is to enable device-independent (DI) information processing. Conditioned on the violation of a Bell inequality, it becomes possible to certify the security or correct functioning of an information processing protocol, without any detailed knowledge of its implementation \cite{Mayers1998,Barrett2005,Acin2006,Acin2007,colbeck_private_2011,Pironio2010,Colbeck2012,Miller2014,Arnon2016}. Prominent examples are quantum key distribution and random number expansion and amplification. In DI settings, it is typically assumed that devices are shielded, i.e., that the experimenters control the inputs which enter into the devices, and that the devices do not leak information on spurious side channels.
 For DI information processing therefore, the minimal non-trivial setting is the Instrumental scenario \figref{fig.DAGs}(b) considered here.

We have shown here that the Bell and Instrumental scenarios are closely related. Though correlations in the simplest Bell scenario, the CHSH scenario, always admit a classical model if they are \textit{directly} projected on the Instrumental scenario, we have shown that their non-classical nature is \textit{entirely} preserved in the Instrumental scenario provided some purely classical local processing is first applied on Alice's side. This finding has important implications: given some non-classical resource $p(ab|xy)$ in an arbitrary Bell scenario, determining whether this resource gives rise to a non-classical behavior in the Instrumental scenario cannot simply be answered by considering the Instrumental  probabilities $p(ab|x)=p(ab|x,y=a)$ (and determining if they are in the classical Instrumental polytope, e.g., using linear programming). Instead, one should also take into account all possible local classical transformations that can be applied to the given non-classical correlation $p$. By failing to consider such trivial, free transformations of a correlation one obtains false negatives from the standard causal inference tools -- correlations appear to be compatible with the classical Instrumental DAG, but actually are not\footnote{Such free transformations are analogous to the concept of \textit{interventions} in the classical causality literature and are known to improve the reliability of causal inferences \cite{Pearlbook}.}. This observation applies to other DAGs derived from Bell-type scenarios, such as the Signalling-Between-Outputs scenario of \figref{fig.DAG_sbo}.

Another outcome of our results is that they lead to Instrumental versions of the CHSH, tilted CHSH, and chained Bell inequalities (see Appendix~\ref{sec:appendix}) that requires fewer input choices than their standard Bell versions {while preserving their fundamental quantum properties.} It is an interesting open question whether there exists Instrumental inequalities which cannot be reduced to standard Bell inequalities via local processing, as the ones we introduced here.

From a fundamental point of view, we have identified a fully device-independent scenario (in particular, one which does not rely on several independent hidden sources \cite{BilocalCorrelations}) that require three random input choices only, whereas the CHSH scenario requires in total four ($2\times 2$) random input choices. We leave it as an open question whether it is possible to find a fully DI scenario where a random choice between \textit{two} values only is sufficient to observe non-classical correlations.

\section*{Note added}
The results presented here partly overlap with those obtained independently in \cite{chaves_quantum_2017}, where Bonet's inequality and the violating quantum and GPT correlations of Section~\ref{se:4} were also introduced, but where the one-to-one relation between Bonet's inequality and the CHSH inequality presented in Section~\ref{se:6} was not noticed. All our results up to Section~\ref{se:6} have been orally presented by S.P. at the Quantum Networks 2017 Workshop, Oxford (UK) in August 2017.

\begin{acknowledgments}
This research was supported by the Foundation Wiener-Anspach, the Interuniversity Attraction Poles program of the Belgian Science Policy Office under the grant IAP P7-35 photonics@be, the Fonds National de la Recherche Scientifique F.R.S.-FNRS (Belgium) under a FRIA grant, and the Perimeter Institute for Theoretical Physics. Research at Perimeter Institute is supported by the Government of Canada through the Department of Innovation, Science and Economic Development Canada and by the Province of Ontario through the Ministry of Research, Innovation and Science.
S.P. is Research Associate of the Fonds de la Recherche Scientifique (F.R.S.-FNRS). T.V.H. acknowledges funding from the F.R.S.-FNRS, through a FRIA grant (Belgium).
ABS acknowledges partial support by the Foundation for Polish Science (IRAP project, ICTQT, contract no. 2018/MAB/5, co-financed by EU within Smart Growth Operational Programme).
\end{acknowledgments}

\setlength{\bibsep}{3pt plus 3pt minus 2pt}
\bibliographystyle{quantum_ph}
\bibliography{trit_scenario}

\appendix

\section{A generalization of the Instrumental scenario and the chained Bell inequalities}\label{sec:appendix}
The Instrumental scenario can be understood as a Bell scenario where Bob's input $y$ is not free, as is usually required in regular Bell experiments,  but entirely determined by Alice's outcome $a$. Naively, it may thus seem surprising that a quantum violation is at all possible in that scenario. However, Bob's input $y$ is not known to Alice \textit{before} outputting her outcome (in which case a classical explanation would always be possible) but \textit{after} it, making a classical simulation a non-trivial task.

\begin{figure}[h]
\centering
\includegraphics[width=0.7\columnwidth]{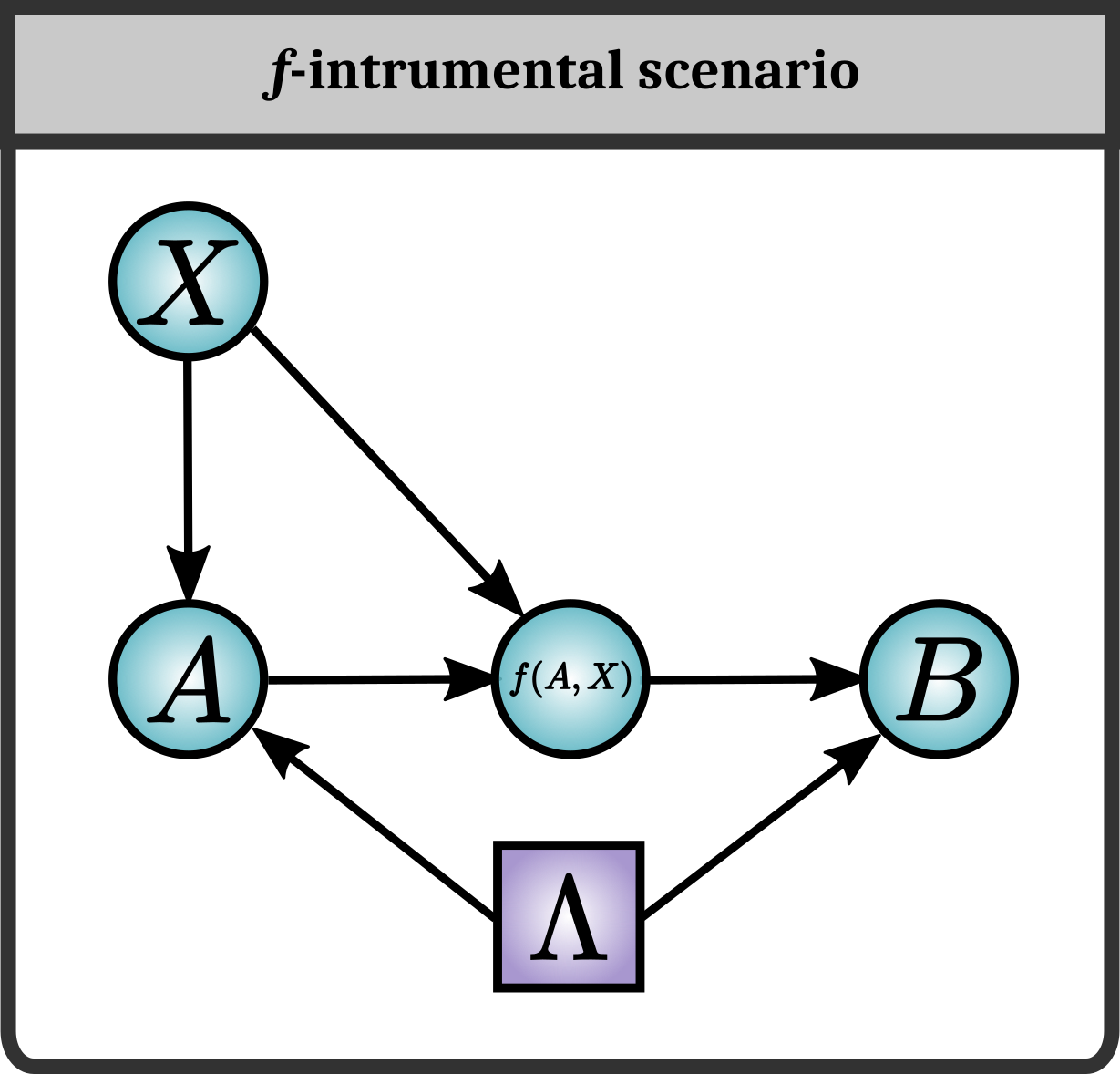}
\caption{DAG for the $f$-Instrumental scenario.}
  \label{fig.finstr}
\end{figure}

In this Appendix, we introduce a slight generalization of the Instrumental scenario where Bob's input $y$ is not merely equal to Alice's output $a$, but is an arbitrary function $y=f(a,x)$ of Alice's output $a$ and input $x$. As in the regular Instrumental scenario, it is thus still the case that Bob's input is entirely determined by the variables on Alice's side. We call this generalized scenario, depicted in \figref{fig.finstr}, the $f$-Instrumental scenario. Provided $f(a,x)$ does genuinely depend on $a$ (so that Bob's input $y$ is not predetermined before Alice's output) and that the knowledge of $f(a,x)$ does not totally reveal the value of $x$ (so that Bob does not know Alice's input $x$ prior to performing his own measurement), one should also obtain generic quantum violations in this new scenario. We show below that this is indeed the case for a specific choice of $f$ which allows us to give a $f$-Instrumental version of the chained Bell inequalities of Braunstein and Caves \cite{braunstein_wringing_1990}.

We consider a $f$-Instrumental scenario  where $X\in\{0,1,\ldots,N\}$ ($N\geq 2$), $A,B\in\{0,1\}$,  and $f(A,X)=(X-A)  \, \mathrm{mod} \, N$. We then define the following chained expression
\begin{equation}
I_N = \sum_{j=1}^{N-1} p(a=b|j) + p(b=0|0) + p(11|N)\,.
\end{equation}
Whenever $N=2$, we recover Bonet's expression (up to a relabeling of inputs and outputs). Using the same technique as in Section~\ref{se:6}, one can show that
\begin{equation}
\textrm{Lifting}\left[ I_{N} \right] = \frac{1}{4} \avg{CH_N}  - p(01|N,N-1)+ \frac{N+1}{2}\, ,
\end{equation}
where
\begin{equation}\label{eq:chained}
\avg{\text{CH}_N}\equiv \sum_{j=1}^{N-1}\avg{A_jB_j}+\avg{A_jB_{j-1}}+\avg{A_0B_0}-\avg{A_0B_{N-1}},
\end{equation}
is the chained Bell inequality \cite{braunstein_wringing_1990}, generalizing the CHSH inequality to the $X,Y\in\{0,1,\ldots,N-1\}$ Bell scenario.
This inequality obeys the following bound
\begin{eqnarray}
\langle \text{CH}_N \rangle \leq \begin{cases}
2N-2 & \text{Classical}\\
2N \cos{\left(\frac{\pi}{2N}\right)}  & \text{Quantum}\\
2N & \text{GPT}
\end{cases}\,,
\end{eqnarray}
translating for the Instrumental version to
\begin{eqnarray}
\langle \text{I}_N \rangle \leq \begin{cases}
N & \text{Classical}\\
N \left[\frac{1}{2} + \frac{1}{2}\cos{\left(\frac{\pi}{2N}\right)} \right]+1/2 & \text{Quantum}\\
N+1/2 & \text{GPT}
\end{cases}\,.
\end{eqnarray}
As before, these values can be achieved from the corresponding strategies for the chained Bell expression for $X\in\{0,1,\ldots,N-1\}$ and using the deterministic assignment $a=1$ when $X=N$.

We can thus interpret the chained Instrumental inequality as a test of the standard chained Bell inequality, but which requires only $N+1$ input choices instead of $N\times N=N^2$ ones.
(Remark that the chained Bell scenario can also be tested using $2N$ input choices if a common referee selects the value $X\in\{0,\ldots,N-1\}$ and then one of the two values $Y\in\{X,X-1\}$, where $X_{-1}=X_{N-1}$, since these are the only inputs appearing in the chained expression (\ref{eq:chained})).

\end{document}